\documentclass[graphicx,prl,aps,preprint,showpacs]{revtex4}
\usepackage{graphicx}

\begin{document}

\title{{\Large Spin injection and perpendicular spin transport\\ in graphite nanostructures}}
\author{T. Banerjee$^{*}$$^{1,2,3}$, W.G. van der Wiel$^{2}$ and R. Jansen$^{3}$}
\affiliation{\vspace*{5mm} $^1$ Physics of Nanodevices, Zernike Institute for Advanced Materials, University of Groningen, The Netherlands\\
$^{2}$ Nanoelectronics Group, MESA$^+$ Institute for Nanotechnology, University of Twente, The Netherlands\\
$^{3}$ MESA$^+$ Institute for Nanotechnology, University of Twente, The Netherlands\\
$^{*}$e-mail: T.Banerjee@rug.nl\\ }


\begin{abstract}
Organic and carbon-based materials are attractive for spintronics
because their small spin-orbit coupling and low hyperfine
interaction is expected to give rise to large spin-relaxation times.
However, the corresponding spin-relaxation length is not necessarily
large when transport is via weakly interacting molecular orbitals.
Here we use graphite as a model system and study spin transport in
the direction perpendicular to the weakly bonded graphene sheets. We
achieve injection of highly (75\%) spin-polarized electrons into
graphite nanostructures of 300-500 nm across and up to 17 nm thick,
and observe transport without any measurable loss of spin
information. Direct visualization of local spin transport in
graphite-based spin-valve sandwiches also shows spatially uniform
and near-unity transmission for electrons at 1.8 eV above the Fermi
level.
\end{abstract}

\maketitle

\clearpage

\indent The efficient injection and transport of spin-polarized
carriers in electronic nanostructures is a subject of intense
research and a basic ingredient of spintronics, a technology in
which digital information is represented by spin
\cite{zutic,chappert}. Organic and carbon-based materials are
attractive for spintronics because spin is only weakly coupled to
other degrees of freedom in these materials, leading to favorably
long spin-relaxation times \cite{naberJPD,coey,dediureview}. Indeed,
successful spin transport through organic materials
\cite{dediu,xiong,moodera1,pramanik,moodera2,cinchetti,drew}, carbon
nanotubes \cite{tsukagoshi,schonenberger,hueso} and graphene
\cite{tombros,suzuki,kawakami} has recently been reported. Yet, the
understanding of injection and transport of spin-polarized carriers
in organic materials is still at its infancy
\cite{coey,dediureview}, and explicit confirmations of very large
spin-relaxation lengths remain scarce. For instance, spin transport
through Alq$_3$ (tris-(8-hydroxyquinoline) aluminium) was reported
to decay with a characteristic length of 10 to 40 nm at low
temperature \cite{xiong,pramanik,drew}, whereas a length scale of 13
nm was observed for amorphous rubrene \cite{moodera2}. The spin-flip
length in the organic semiconductor CuPc (copper phthalocyanine) was
found to be 10 to 30 nm \cite{cinchetti}. For comparison, the
spin-diffusion length in silicon was recently determined to be 200
to 300 nm at room temperature \cite{dash}, and larger values
have been obtained at room temperature in graphene \cite{tombros}.\\
\indent Although a large spin {\em lifetime} is desirable, the
corresponding spin-relaxation {\em length} is not always large,
because it is also determined by transport parameters (such as
the carrier mobility and the diffusion constant). These are
equally important. Transport in most organic compounds is
complicated by the rather localized nature of the electronic
states derived from weakly interacting molecular orbitals.
Consequently the description of spin diffusion and relaxation for
hopping-like conduction is under debate
\cite{coey,dediureview,bobbert}. On the other hand, carbon
nanotubes and graphene exhibit band conduction with high carrier
mobility. Graphite, with its anisotropic resistivity
\cite{anisotropy}, forms an interesting system that combines the
different transport regimes. The conductivity is large within the
plane of the graphene sheets enabled by highly delocalized
electronic states. However, in the direction perpendicular to the
carbon sheets the $\pi$ orbital overlap is limited, and the
conductivity and mobility are at least two orders of magnitude
smaller. Transport in this direction thus resembles that found in
many organic materials, and hence graphite is a unique system to
study spin transport across interfaces with weak electronic
interactions.\\
\indent Our experiment thus involves a spin-valve sandwich
consisting of two ferromagnetic metal layers and a graphite
spacer. The first ferromagnet acts as a spin filter, producing a
spin-polarized current that is subsequently injected into the
graphite. After transmission of the graphite, the electrons
proceed into the second ferromagnet that acts as analyzer of the
transmitted spin polarization. When spin is conserved in the
graphite spacer, the total transmission is largest when the
magnetization of the two ferromagnets is aligned parallel (P),
and smaller for antiparallel (AP) alignment. One obtains a
magnetocurrent MC$=(I^P-I^{AP})/I^{AP}$, where I$^P$ and I$^{AP}$
denote the transmitted current for the P and AP magnetic state,
respectively. Spin relaxation in the graphite spacer tends to
equalize I$^P$ and I$^{AP}$ and reduces the MC. Comparing the MC
for structures with different graphite thickness thus provides
information on the spin relaxation in the graphite spacer.\\
\indent We employ Ballistic Electron Magnetic Microscopy
\cite{rippard1,rippard2,haq} (BEMM, see Fig. 1a), which is uniquely
suited to study perpendicular spin transport through buried layers
and their interfaces \cite{beem,beemreview}. The technique is based
on scanning tunneling microscopy (STM) and provides direct
visualization of any nanoscale spatial inhomogeneity of the
transport, which has always been an issue for the interpretation of
magnetoresistive measurements particularly for spin valves with
organic spacers \cite{dediureview}. Another feature of BEMM is that
the electron energy can be varied, typically between 0.3 and 2 eV,
giving valuable information on (spin-) transport and fundamental
excitations not accessible by ordinary conduction at the Fermi
energy. We use BEMM to demonstrate perfect transmission of
spin-polarized electrons perpendicularly through graphite
nanostructures of up to 17 nm thick, corresponding to 51 sheets of
graphene. The graphite nanoflakes are prepared by sonication of
exfoliated flakes of HOPG (SPI-2 grade, density 2.27 g/cm$^{-3}$,
resistivity of 4$\times$10$^{-5}$ $\Omega$cm within the plane, and
1.5$\times$10$^{-1}$ $\Omega$cm perpendicular to the plane) in
VLSI-grade isopropyl alcohol. Characterization by atomic force
microscopy (see Fig.1b and c) shows that the flakes are typically
around 10 nm in height with a few being as high as 20 nm. Their
lateral dimension is between 100 and 500 nm. The graphite flakes are
randomly distributed over the surface, where an area of 5$\times$5
$\mu$m$^2$ on average contains a few nanoflakes that can be located
without extensive searching.
\\
\indent For the BEMM experiments, we start with an n-type Si(100)
substrate (resistivity 5-10 $\Omega$cm) having a 300 nm thick
SiO$_{2}$ with circular contact holes of 150 $\mu$m diameter. After
a final etch in HF acid to remove any native oxide, metal layers
were deposited by thermal evaporation using a molecular beam epitaxy
system (base pressure 10$^{-10}$ mbar). First, a 8 nm Au layer was
evaporated to form a Au/Si Schottky barrier, followed by 3 nm of
Ni$_{80}$Fe$_{20}$ and 3 nm of Au, the latter providing a chemically
inert cap layer. The sample was then taken out of the deposition
chamber for {\em ex situ} transfer of the graphite nanoflakes. Using
a micro-syringe, the nanoparticle solution is dispersed onto the
Si/Au/Ni$_{80}$Fe$_{20}$/Au template and the solvent is allowed to
dry, leaving graphite nanoflakes behind. The sample was then
re-introduced into the deposition system and a stack of
Au(3nm)/Co(3nm)/Au(4nm) was
evaporated.\\
\indent The final structure consists of regions without any
graphite, and regions with a graphite nanoflake sandwiched
between two ferromagnetic layers via intermediate layers of Au
(see Fig. 1a). The STM tip that is used to inject current into
the structure can then be positioned in a location with or
without graphite (location (1) and (2), respectively). The
resulting transmission and MC can thus be compared directly. For
all measurements, the metal surface of the sample is grounded and
negative voltage V$_T$ is applied to the STM tip with the tunnel
current I$_T$ kept constant using feedback. The energy of the
injected electrons is given by eV$_T$ and transport in the
metal/graphite sandwich is thus by hot electrons
\cite{jansenreview}. Before reaching the graphite, the electrons
are spin filtered in the Co metal layer that preferentially
transmits hot electrons of majority spin due to spin-dependent
scattering \cite{jansenreview,vlutters}. A 3 nm thick Co film is
known to transmit hot electrons that have 75\% spin polarization
\cite{jansenreview,vlutters}. After injection and transport
through the graphite and spin analysis in the second
ferromagnetic layer, the transmitted electrons are collected in
the conduction band of the n-type Si substrate having a separate
(third) electrical contact. Collection in the Si is possible only
\cite{beem,beemreview} for those electrons that have retained
sufficient energy and the proper momentum to cross the 0.8 V high
Schottky barrier at the Au/Si interface, making the collected
current I$_C$ sensitive to scattering during transport in the
graphite sandwich. All BEMM measurements were performed at 150 K
using PtIr metal tips. Details of the BEMM setup have been
described elsewhere \cite{haq,HaqAPL05,BanerjeePRL}.\\
\indent The inset of Fig. 2a shows a conventional topographic STM
image. The location of the approximately circular graphite nanoflake
can clearly be identified and the granular morphology of the Co and
Au layers on top of the graphite and besides it can be seen. The
graphite flake was 17 nm in height. The STM tip was positioned at
the centre of the graphite flake, and the transmitted current was
measured as a function of $V_T$ for P and AP alignment of the Co and
Ni$_{80}$Fe$_{20}$ magnetization (Fig. 2a). As expected
\cite{rippard2,haq,beem,beemreview}, the transmission is nonzero
only for V$_T<-0.8$ V when the electron energy is above the Au-Si
Schottky barrier height. Most notably, the current is largest for
the P state and more than a factor of three smaller for the AP
configuration. The large difference between I$_C^P$ and I$_C^{AP}$
with the graphite as spacer demonstrates efficient transmission of
spin polarization through the graphite. This was further proven by
measuring, with the STM tip still above the centre of the graphite
nanoflake, the transmitted current while sweeping the magnetic field
through a complete cycle from $+$100 Oe to $-$100 Oe and back (Fig.
2b, taken with constant tunnel current (3 nA) and tip bias ($-$1.8
V)). The magnetization of both ferromagnets was first saturated in a
magnetic field of $+$100 Oe to obtain a P state yielding largest
transmission (0.7-0.8 pA). When the magnetic field is swept to
negative values, a transition to the AP state with lower
transmission (0.2-0.3 pA) occurs due to magnetization reversal of
the soft Ni$_{80}$Fe$_{20}$ layer, followed by a transition back to
the P state around $-$40 Oe when the Co magnetization is also
reversed. A similar behaviour is observed on the retrace, with
hysteresis. The corresponding MC is 250$\pm$30\%, demonstrating that
a significant spin polarization is injected into the graphite and
transmitted perpendicularly through the graphite spacer and
its interfaces with the metals.\\
\indent While the above results establish spin transport in
graphite, a reliable quantitative analysis exploits the imaging
capability of BEMM. This is required because of possible local
variations of the transmitted current and the MC. The left two
panels of Fig. 3 show a 0.6$\times$0.6 $\mu$m$^{2}$ STM topography
image and the corresponding height profile of the graphite
nanoflake, which in this case is about 7 nm thick. The centre and
right images are spatial maps of the transmitted current I$_C$ for P
and nominally AP state, respectively, all taken in the same area
(V$_T=-$1.8 V). For the P state the transmitted current is largest
and equal to about 0.25 pA per nA of injected tunnel current. For
the image on the right, the current is strongly reduced in most of
the area due to the AP alignment, and magnetic domain contrast
appears in the left part, as often observed in BEMM on metal
spin-valve stacks \cite{rippard1,HaqAPL05}. However, the most
important feature is that there is no significant difference in the
transmission in the area with the graphite flake, as compared to the
surrounding area without graphite. This applies to the P and to the
AP state, as can be seen in the cross sections taken along a line
intersecting the graphite flake (bottom panels). This leads to the
following main quantitative conclusions: (i) the transmission
through the graphite is nearly perfect, i.e., there is no
significant attenuation of the current, and (ii), the MC with and
without the graphite is identical, implying that the electrons are
transmitted through the graphite and its interfaces
without any measurable loss of spin polarization.\\
\indent A similar set of data was obtained on the thicker graphite
flake of 17 nm (Fig. 4, also for V$_T=-$1.8 V). Again, the
transmission for P and AP state occurs without significant
attenuation due to the graphite, and there is no reduction of the MC
in the area where the additional graphite spacer is present (note
that there are some small inhomogeneities at the edges of the
graphite flake, causing strong attenuation of the hot-electron
transmission for the P as well as the AP state (dark spots in the
spatial maps of I$_C$)). More precise analysis is done using the
distribution of current values across the images of Fig. 3 and 4
containing the 7 nm and the 17 nm thick graphite. The resulting
histograms of the transmitted current are displayed in Fig. 5a and
5b, where the histograms in blue are obtained from the area with the
graphite, while the red histograms correspond to the surrounding
area without graphite. To first order, the histograms with and
without graphite overlap, as expected from the images and cross
sections already described. More precisely, the mean values of the
transmitted current on the 7 nm flake are $I_C^P=0.24\pm0.02$ pA/nA
and $I_C^{AP}=0.13\pm0.02$ pA/nA, whereas away from the graphite we
have $I_C^P=0.25\pm0.01$ pA/nA and $I_C^{AP}=0.14\pm0.01$ pA/nA.
Similarly, for the 17 nm graphite flake we have $I_C^P=0.26\pm0.02$
pA/nA and $I_C^{AP}=0.12\pm0.02$ pA/nA, and surrounding the flake we
have
$I_C^P=0.255\pm0.01$ pA/nA and $I_C^{AP}=0.12\pm0.01$ pA/nA.\\
\indent As previously established \cite{jansenreview}, for hot
electrons the current transmitted through a
ferromagnet/spacer/ferromagnet stack can be described as a
product of the transmissions of each layer if spin is conserved:
\begin{eqnarray}
I_C^{P} \propto T^M_{NiFe}T_{Gr}T^M_{Co} +
T^m_{NiFe}T_{Gr}T^m_{Co}, \label{eq1}\\
I_C^{AP} \propto T^M_{NiFe}T_{Gr}T^m_{Co} +
T^m_{NiFe}T_{Gr}T^M_{Co}. \label{eq2},
\end{eqnarray}
where $T^M$ and $T^m$ denote the transmission of hot electrons of
majority (M) and minority (m) spin in the ferromagnetic layers, and
$T_{Gr}$ is the transmission of the graphite (not spin dependent).
All transmission factors depend exponentially on the layer thickness
\cite{jansenreview,vlutters}. Specifically, we have
$T_{Gr}\,\propto\,e^{-d/\lambda_{k,E}}$, where d is the thickness of
the graphite spacer and $\lambda_{k,E}$ is the length scale
associated with the current attenuation due to {\em spin-conserving}
scattering processes that change the energy or momentum of the hot
electrons (not to be confused with the spin-relaxation length).
Given that the transmission of the graphite is near unity (Fig. 5),
we can conclude that the value of $\lambda_{k,E}$ must be at least
on order of magnitude larger than the graphite thickness used.
Hence, $\lambda_{k,E}$ is conservatively estimated to be larger than
100 nm at 1.8 eV above the Fermi level. In materials where spin
relaxation is dominated by scattering involving the spin-orbit
interaction (Elliott-Yafet mechanism), there is an approximate
scaling \cite{zutic,soscaling} between the momentum scattering time
$\tau$ and the spin-relaxation time $\tau_s$. The ratio
$\tau/\tau_s$ depends on the spin-orbit interaction, and for
materials with light elements (such as carbon) and weak spin-orbit
interaction, we have $\tau/\tau_s<<1$, implying that many scattering
events are needed to create a significant change of spin. Hence, the
spin-relaxation length should be much larger than $\lambda_{k,E}$,
and may thus approach the micron range. This is a rather surprising
result for transport in the direction perpendicular to the graphene
sheets, which are coupled in this direction by $\pi$ orbitals with
limited overlap. A much shorter spin scattering length was therefore
expected, in analogy with spin transport via weakly interacting
orbitals in organic materials, which yields spin-scattering lengths
in the 10 to 40 nm range
\cite{xiong,pramanik,moodera2,cinchetti,drew}. Nevertheless, the
spin-flip length for perpendicular transport in graphite is found to
be significantly larger than the graphite thickness of up to 17 nm
used here, and spin is thus essentially conserved. Note that
transport parameters and spin-scattering lengths for hot electrons,
as used here, are not the same as those of electrons at the Fermi
energy. For hot electrons the carrier velocity is different
(generally higher) compared to that of Fermi electrons, while the
scattering cross section is also different
(generally larger because of the larger phase space for
elastic scattering and the additional inelastic scattering channels that are available for hot electrons).\\
\indent Graphite can be obtained with high purity and hence is an
ideal model system for a detailed investigation of spin transport
across interfaces with weak electronic interactions, without the
complications of impurities that are often present in organic
compounds \cite{coey,dediureview}. This offers hope for a meaningful
comparison with theoretical descriptions, for which our results
provide a challenging benchmark. Also of particular interest in this
regard is the recent theoretical prediction of strong spin filtering
at crystalline interfaces between graphene/graphite and ferromagnets
in perpendicular transport geometry \cite{karpan,karpanprb}.
Combined with the perfect transmission of spin-polarized electrons
through graphite, even at an energy of 1.8 eV, as demonstrated here,
this raises prospects for graphite as a potential material for
spintronics devices. The results also highlight the unique
capability of our scanning-probe-based technique to study and
directly visualize local spin transport in organic and carbon-based
materials at
the nanoscale.\\

\indent {\bf Acknowledgement.} We are grateful to Prof. P.J.
Kelly and his team members for sharing their transport
calculations, to Dr. R. Salvio for useful discussion on the
sonication process, and to Prof. B.J. van Wees for his critical
reading of the manuscript. This work was financially supported by
the NWO-VIDI program and the Netherlands Nanotechnology Network
NANONED (supported by the Ministry of Economic Affairs).\\

\clearpage

\begin{figure}[htb]
\hspace*{-5mm}\includegraphics*[width=90mm]{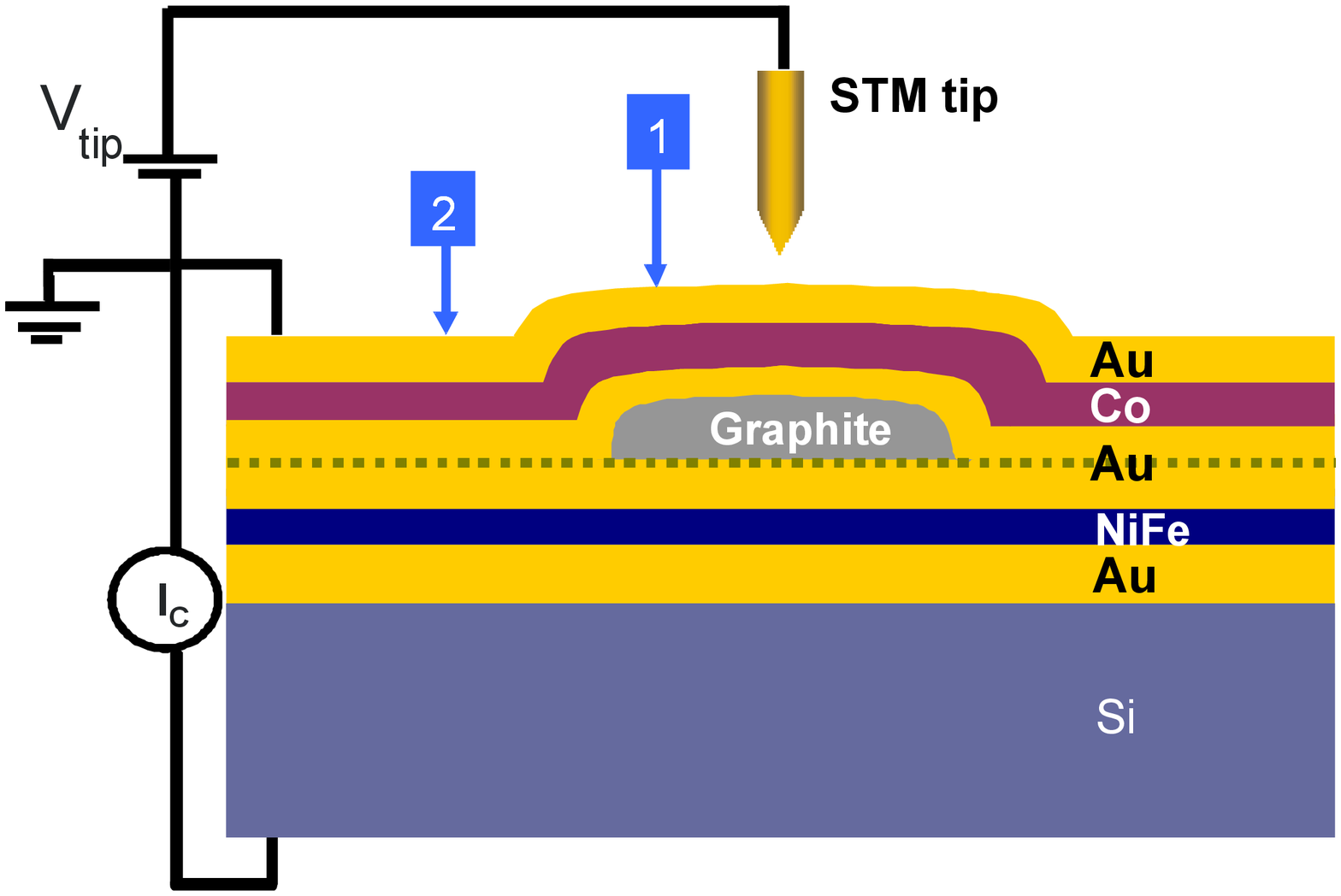}
\hspace*{5mm}\includegraphics*[width=60mm]{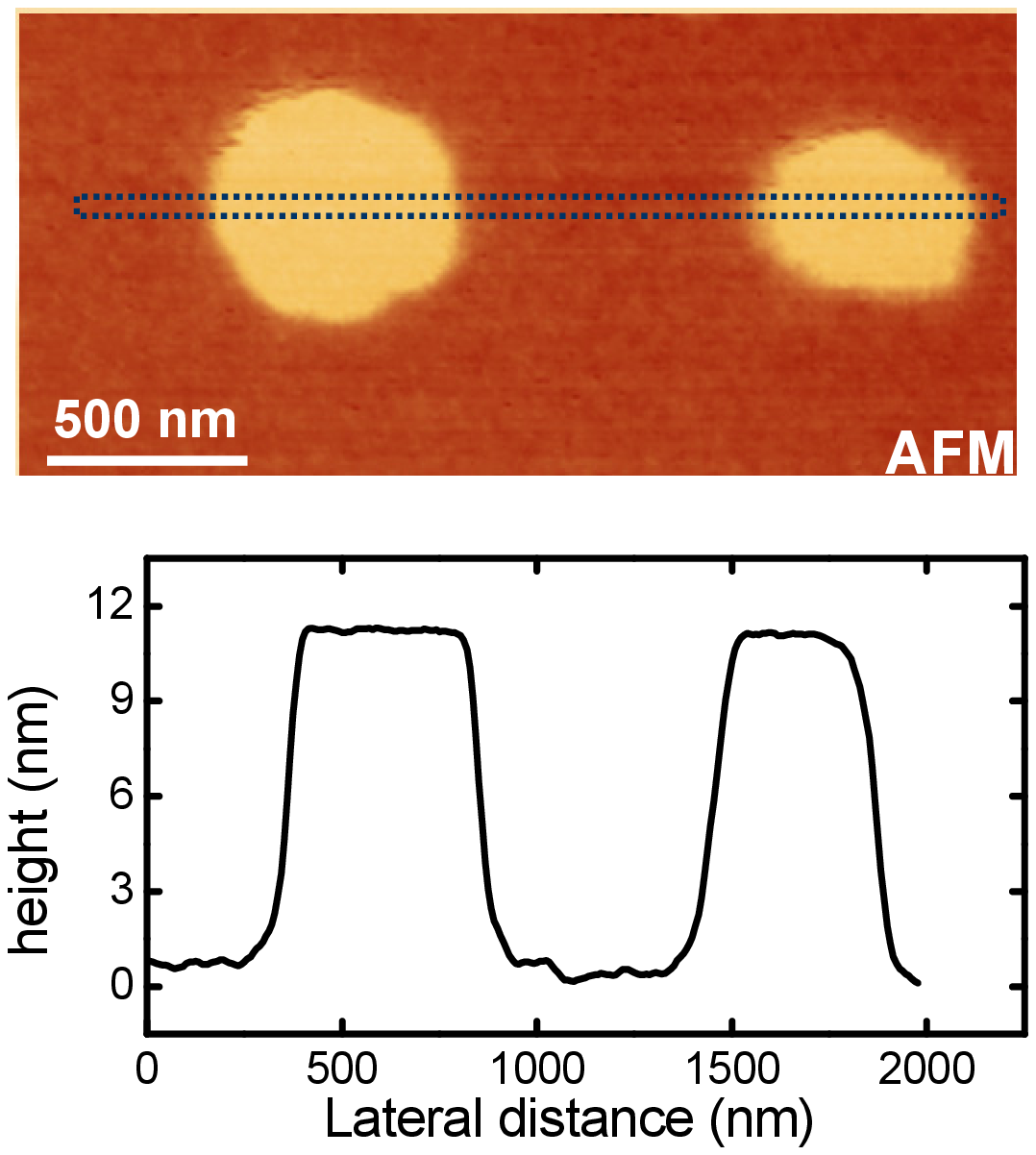}
\vspace*{3mm} \caption{(a) Schematics of the BEMM technique. The
sample consists of a Si substrate coated with a layer stack of
Au(8nm)/Ni$_{80}$Fe$_{20}$(3nm)/spacer/Co(3nm)/Au(4nm), where the
spacer is either a graphite nanoflake sandwiched between two Au
layers of 3 nm (location 1), or just 3+3 nm Au (location 2). The
STM tip is used to locally inject electrons into the sample by
tunneling at bias voltage V$_T$ between tip and Au surface. The
current I$_{C}$ transmitted perpendicularly through the stack is
collected in the Si with a third electrical contact. (b) AFM image
of a complete sample structure (including the top metal coating),
showing two graphite nanoparticles of approximately circular
shape. (c) Height profile along the horizontal line intersecting
both graphite flakes, as indicated in the image.} \label{fig1}
\end{figure}

\vspace*{5mm}

\begin{figure}[htb]
\hspace*{5mm}\includegraphics*[width=80mm]{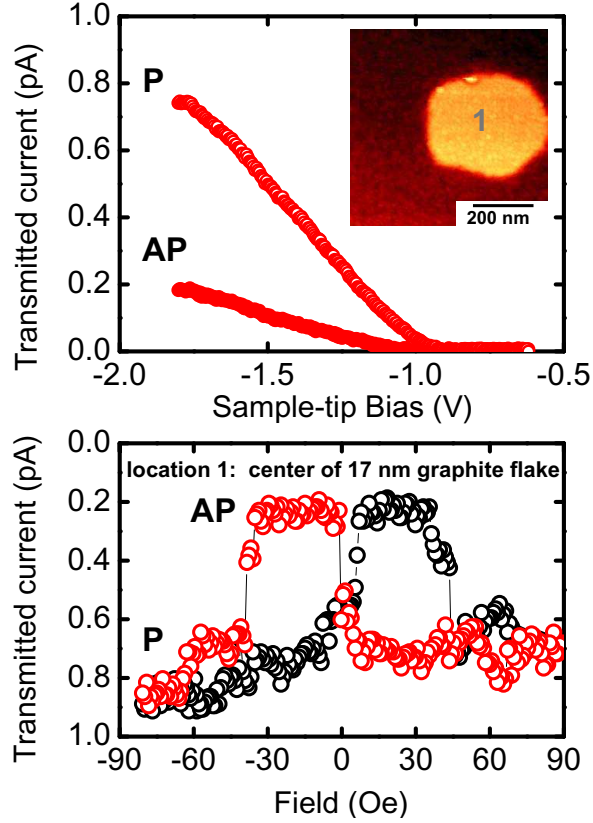}
\caption{(a) I$_{C}$ versus V$_T$ with the tip positioned at the
centre of the 17 nm thick graphite flake (location 1). We first
applied an in-plane magnetic field of $-$100 Oe, then reversed the
field to $+$20 Oe (spectrum for AP state), and then increased the
field to $+$100 Oe (spectrum for P magnetic state). The inset
shows the 0.6$\times$0.6 $\mu$m$^{2}$ topographic STM image (taken
at V$_T=-$1V and I$_T=$0.6 nA) with location 1 indicated. (b)
Local spin-valve measurement at the centre of the graphite flake
(location 1), showing I$_{C}$ versus magnetic field swept from
negative to positive (black) or vice versa (red), at constant
V$_T=-$1.8V and I$_T=$3 nA. T = 150 K.} \label{fig2}
\end{figure}

\vspace*{5mm}

\begin{figure}[htb]
\centering\includegraphics*[width=160mm]{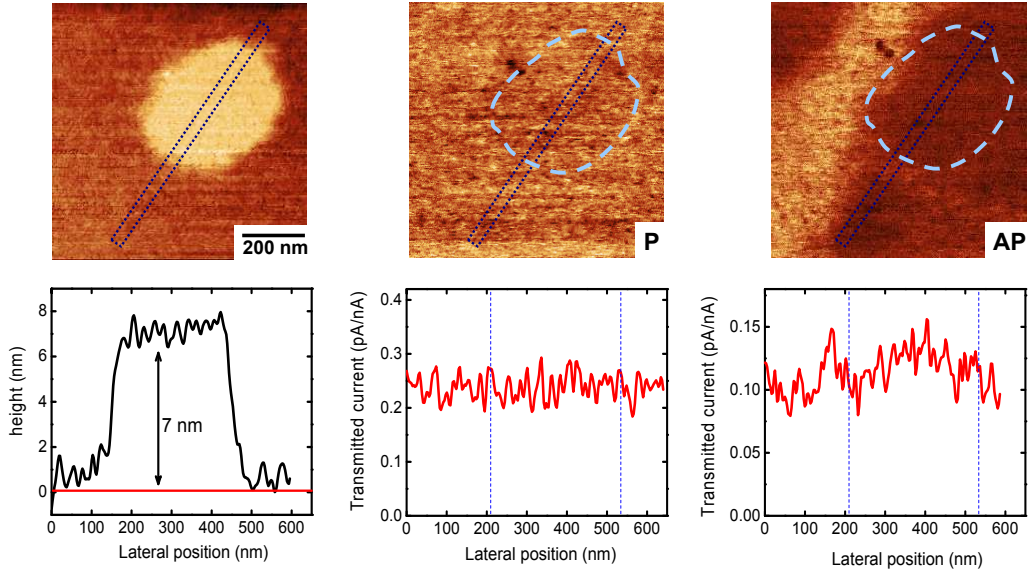}
\vspace*{0mm}\caption{Spatially resolved spin transmission through
graphite spin-valve sandwiches. (a) Topographic STM image. (b)
corresponding BEMM image of transmitted current I$_C$ for $+$100
Oe magnetic field (P magnetic state). (c) BEMM image of I$_C$ for
nominally AP magnetic state, obtained by first setting the
magnetic field to $-$100 Oe and then reversing it to $+$20 Oe. In
the BEMM images, dark (bright) areas have low (high) transmitted
current. (d,e,f) cross-section profiles of the signals along the
line intersecting the graphite, as indicated in the images. Dashed
lines in the cross-sections denote the boundary of the graphite
flake. All data was obtained at V$_T=-$1.8V and I$_T=$3 nA in the
same 0.6$\times$0.6 $\mu$m$^{2}$ area containing a 7 nm high
graphite flake. The rest of the layer stack is identical to that
of Fig. 1. The transmitted current is given in pA/nA, as it is
normalized to I$_T$. T = 150 K.} \label{fig3}
\end{figure}

\vspace*{5mm}

\begin{figure}[htb]
\centering\includegraphics*[width=160mm]{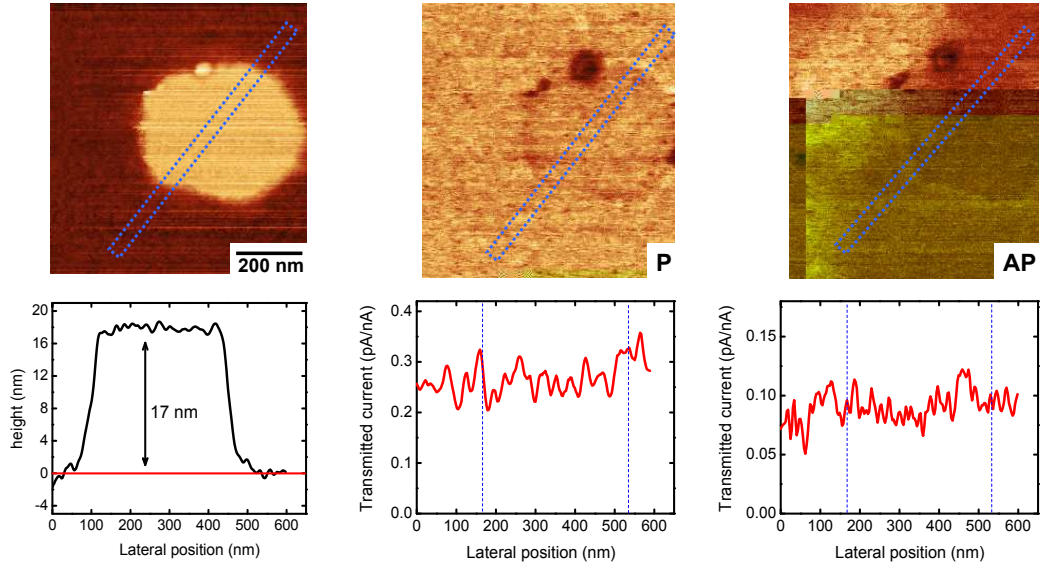}
\vspace*{0mm}\caption{Spatially resolved spin transmission through
17 nm of graphite. Similar set of data as shown in Fig. 3,
obtained with exactly the same parameters, but in a region
containing a 17 nm thick graphite flake (same flake as in Fig.
2(a),(b)).} \label{fig4}
\end{figure}

\vspace*{5mm}

\begin{figure}[htb]
\centering\includegraphics*[width=160mm]{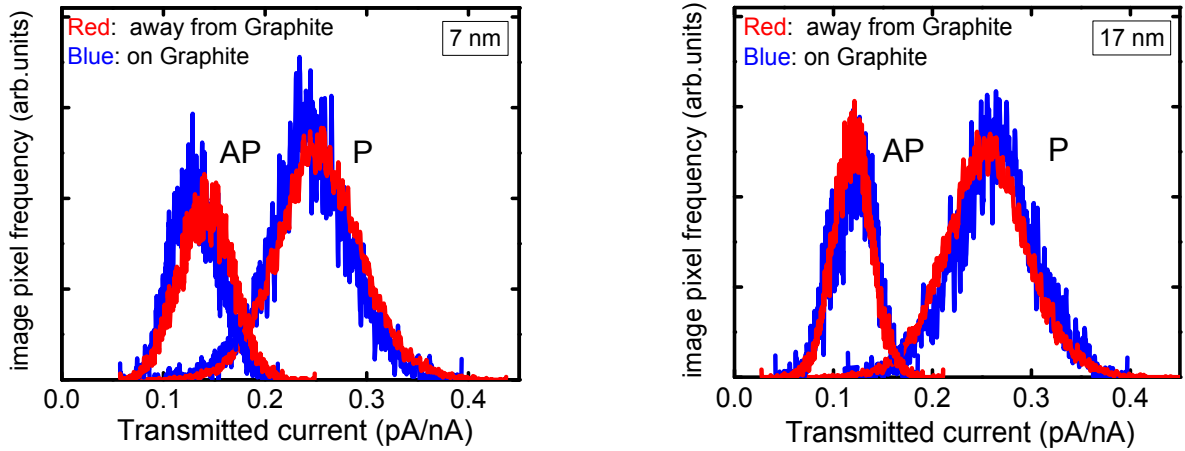}
\caption{Quantitative analysis using histograms of current
distribution. (a) Histograms of the distribution of transmitted
current for P and AP state, derived from BEMM images of Fig. 3
containing the 7 nm graphite flake. Histograms in blue are
obtained from a square section of the image located fully within
the boundary of the graphite flake, while histograms in red are
obtained from an area not containing the graphite.  The
transmitted current is given in pA/nA, as it is normalized to
I$_T$. Note that for the AP state, the left part of the images
(containing magnetic domain contrast) was excluded from the
analysis. (b) Similar set of histograms, but now derived from BEMM
images of Fig. 4 containing the 17 nm graphite flake.}
\label{fig5}
\end{figure}

\end{document}